# Prototype of a multi-host type DAQ front-end system for RI-beam experiments

Hidetada Baba for the RIBFDAQ Collaboration

***The multi-host type DAQ front-end system is proposed and a prototype system is developed. In general, CAMAC/VME type ADC modules have a single trigger-input (or gate-input) port. In contrast, a prototype of a new system has multiple trigger-input ports. In addition, the Wilkinson-type and successive approximation ADCs have the dead time, whereas, this prototype system utilizes the combination of Flash-ADC and FPGA that enabling the dead-time free system. Corresponding to the trigger-input ports, data are sent to different back-end systems. So, a legacy ADC module is a 1-to-1 system, but, this proposing system is a 1-to-X system without loss. This system will be applied for nuclear physics experiments at RIKEN RIBF which produces intense RI-beams. This multi-host type DAQ front-end system enables us to perform different experiments simultaneously at the same beam line. In this contribution, the concept and the performance-test results of the front-end system is shown.***

## I. Introduction

RIKEN Radioactive Isotope Beam Factory (RIBF) is an accelerator facility producing RI beams in Japan [1]. In RI-beam facilities, an experimental group occupies the whole beam line to measure specific physics events. However, the cross-section of nuclear reactions is typically much less than 1 barn (< $10^{-28}$ m2). Therefore, a large number of RI-beams are not reacted. It is possible that such non-reacted RI-beams could be reused by installing multiple experimental setups at different places. Actually, RIBF has a long beam line and there are many detector sections that can be placed user specific systems. If two different experiments are carried out simultaneously at different detector sections in RIBF, RI-beams have to be shared. In this case, difficulty is the RI-beam identification. RI-beams should be identified by using beam line detector information on an event-by-event basis because different nuclides come one after the other. Due to the dead time of the current ADC/TDC front-end system (~200 us/event), the rate of event triggers should be below a few kHz. If two different event triggers from different experimental setups are applied randomly, the dead time of the front-end system will be much worse. In addition, the current front-end system can send online data to only one destination, i.e. 1-to-1 system. Here, a new multi-host type DAQ front-end system is proposed for RI-beam experiments. The concept of this system is shown in Figure 1.

Hidetada Baba (e-mail: baba@ribf.riken.jp), with RIKEN Nishina Center, Hirosawa, Wako, Saitama 351-0198, Japan.

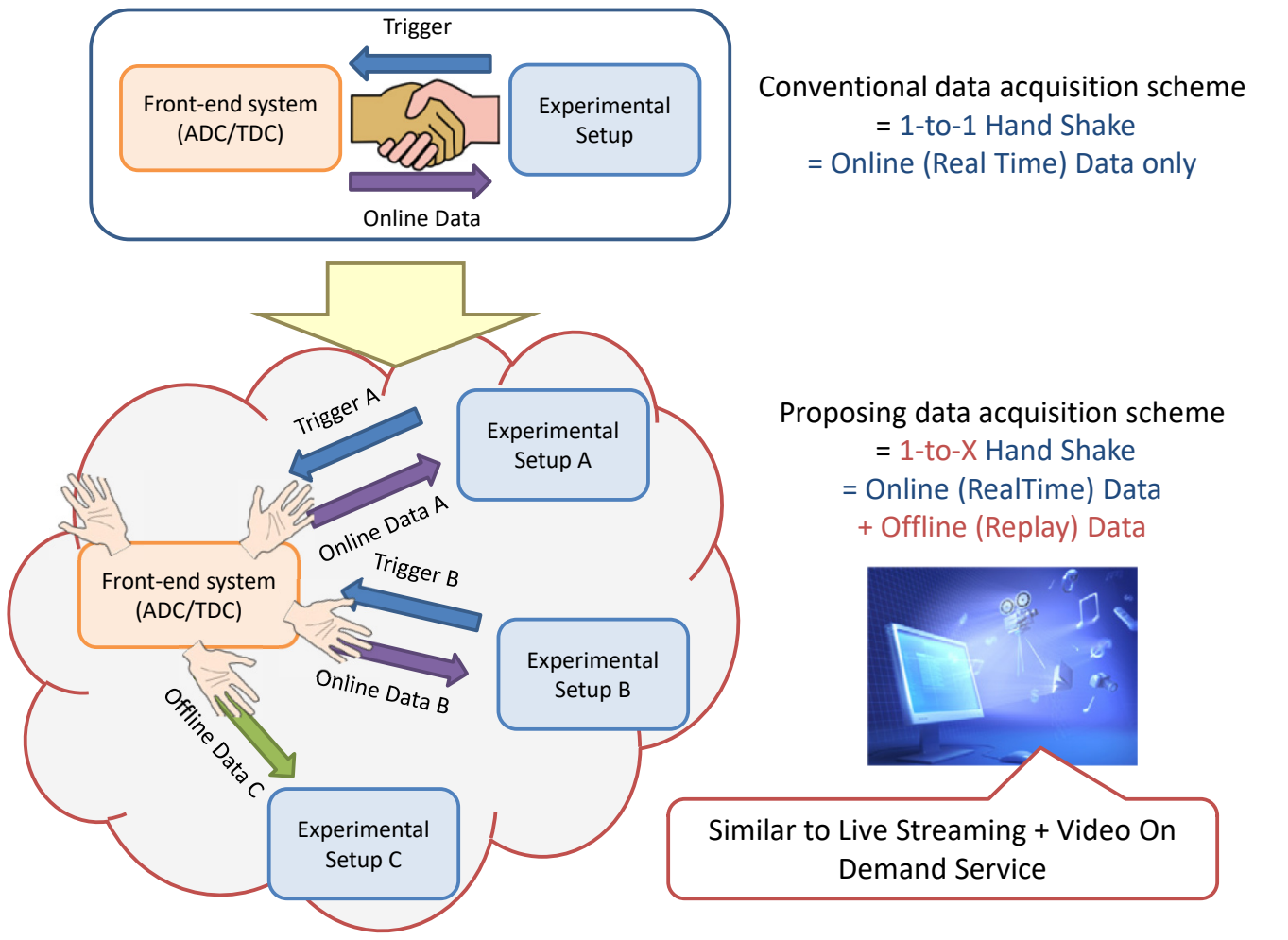

Fig. 1. Concept of the multi-host DAQ front-end system.

## II. Design

This research project aims at demonstrating the new data acquisition scheme for RI-beam experiments. In RIKEN RIBF, there is a long beam line (Figure 2). RI-beams are analyzed by detectors in the BigRIPS section [2]. This BigRIPS section is common for all experiments. There are several user detector sections. By developing the new multi-host type DAQ front-end system, individual experiments at different user detector sections can work simultaneously.

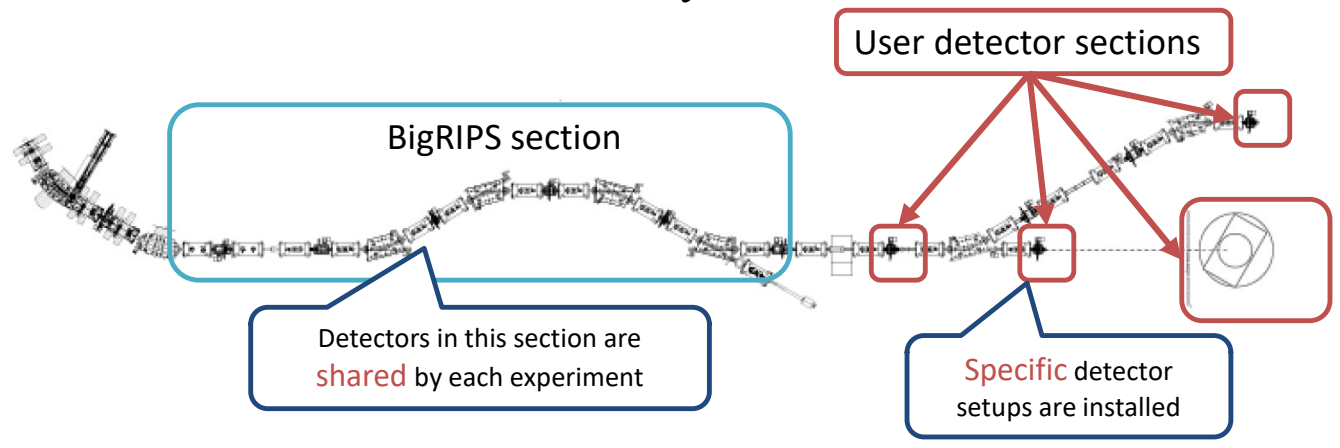

Fig. 2. A part of RIKEN RIBF beam line.

Current data acquisition scheme in RIBF is shown in Figure 3. The details of DAQ in RIBF is described in Ref.[3].The user trigger is applied to front-end systems for BigRIPS and User sections. Data are transferred to the event builder (computer program), and merged online. Also, merged data are stored to the disk. To introduce 1-to-X system, the shared front-end system (BigRIPS section) should work without dead time. In this project, the development of the multi-host type DAQ front-end system is the most important part. Figure 4 shows proposing data acquisition scheme in RIBF. The multi-host type DAQ front-end system accepts multiple trigger signals from different user sections (User trigger A and B in Figure 4). According to the trigger source, the front-end system distributes data to the user event builder. Basically, this system sends data to the event builder in online (RealTime mode), at the same time, all front-end's data will be stored in the disk together with time-stamp information. Time-stamp information corresponds to the absolute date-time. In RIBF, the time-stamping system has been introduced and it can synchronize the time information with 10 ns precision between different front-end systems. In case of measurements of the life time of RI-beams, the time scale is a millisecond order. Therefore, data are separately taken, and after the measurement, these data are merged offline (Replay mode). User Section C in Figure 4 shows this offline scheme.

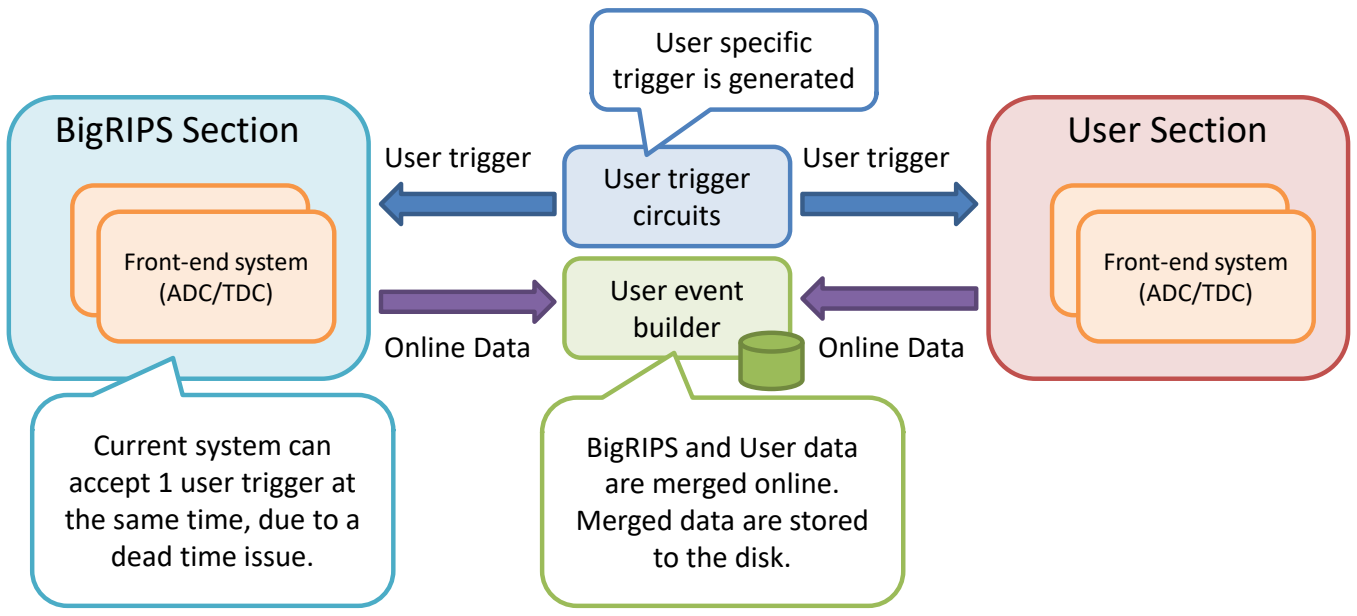

Fig. 3. Current data acquisition scheme in RIBF.

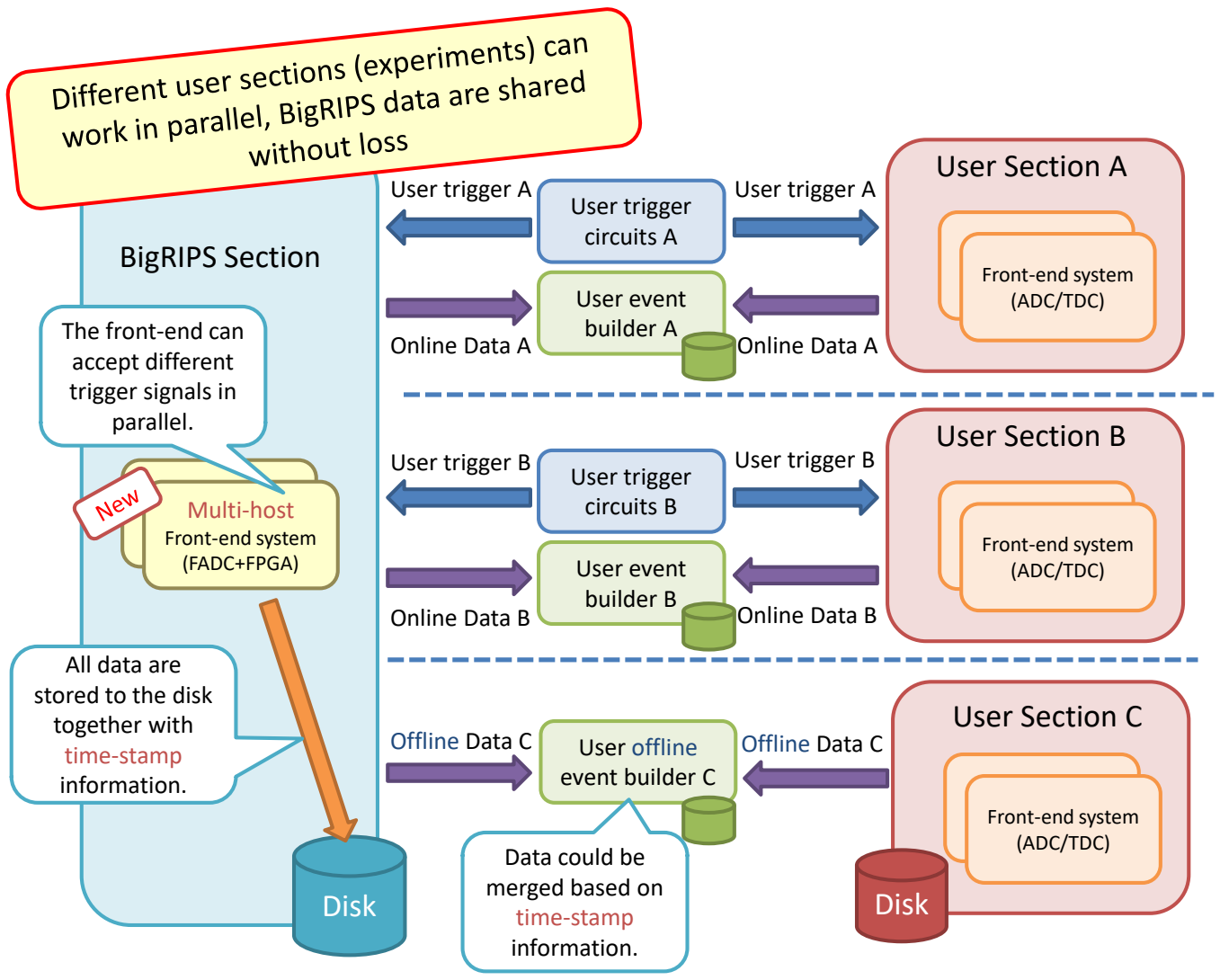

Fig. 4. Proposing data acquisition scheme in RIBF.

## III. PROTOTYPE HARDWARE AND TEST RESULTS

The prototype of the multi-host type DAQ front-end system is developed based on the Cosmo-Z board (Figure 5) which is distributed from TokushuDenshiKairo Inc. [4]. This board contains FlashADC and Xilinx Zynq-7000 SoC [5]. Zynq-7000 SoC is a new device that has ARM processor together with the field programmable gate array. Detector signal is analyzed by the gate array part in real time, and converted data are managed by the processor area (with Linux OS). The Cosmo-Z board has 8 channel of 12-bit 125-MHz FlashADC (Analog Devices AD9633-125). In this contribution, the sub board which has 2 channel of 16-bit 1-GHz FlashADC (Texas Instruments ADS54J60) is also tested.

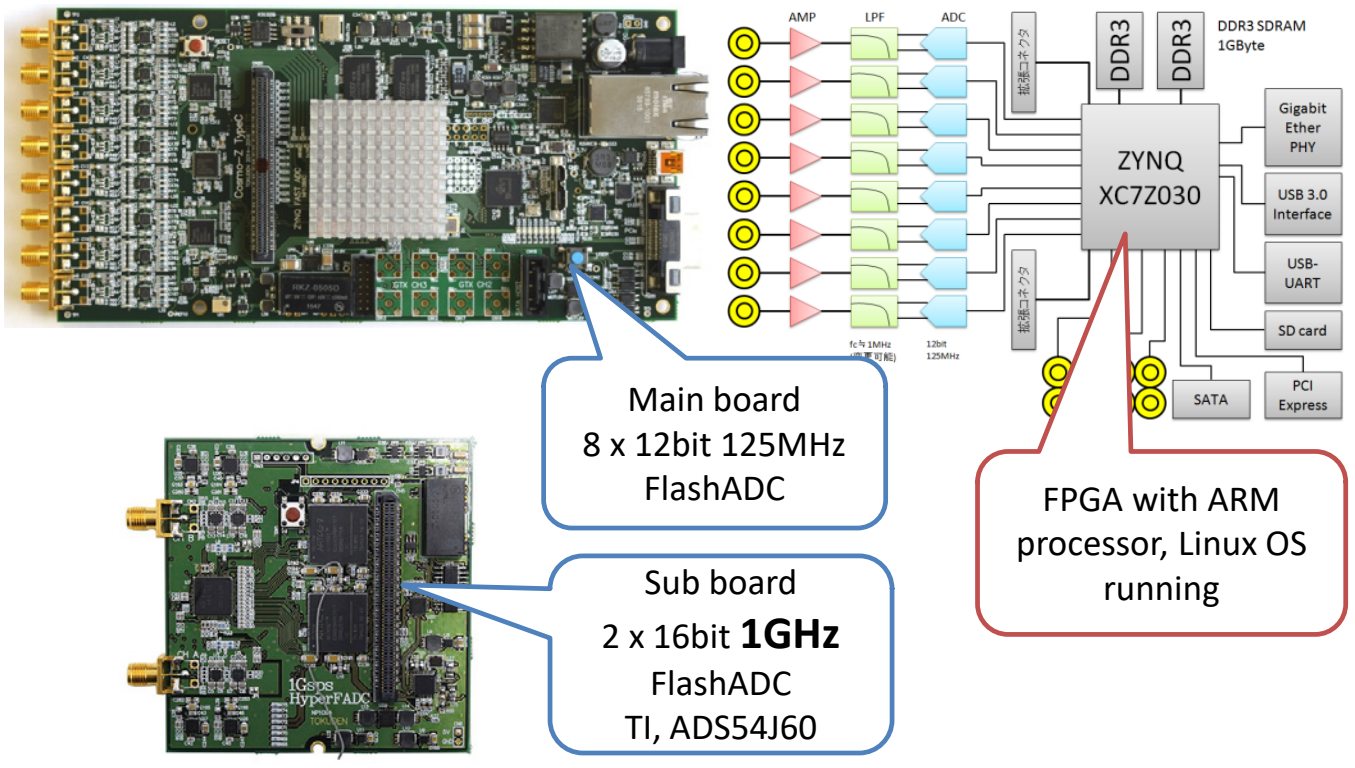

Fig. 5. Photo and schematic of the Cosmo-Z board and sub board.

The effective number of bits (ENOB) of 125-MHz and 1-GHz FlashADCs are measured. Results are shown in Figure 6 and 7. ENOB of 125-MHz (1-GHz) Flash ADC was 11.8 (11.3) bit. These values are close to its data sheet specifications (12 bit for 125-MHz and 11.8 for 1-GHz).

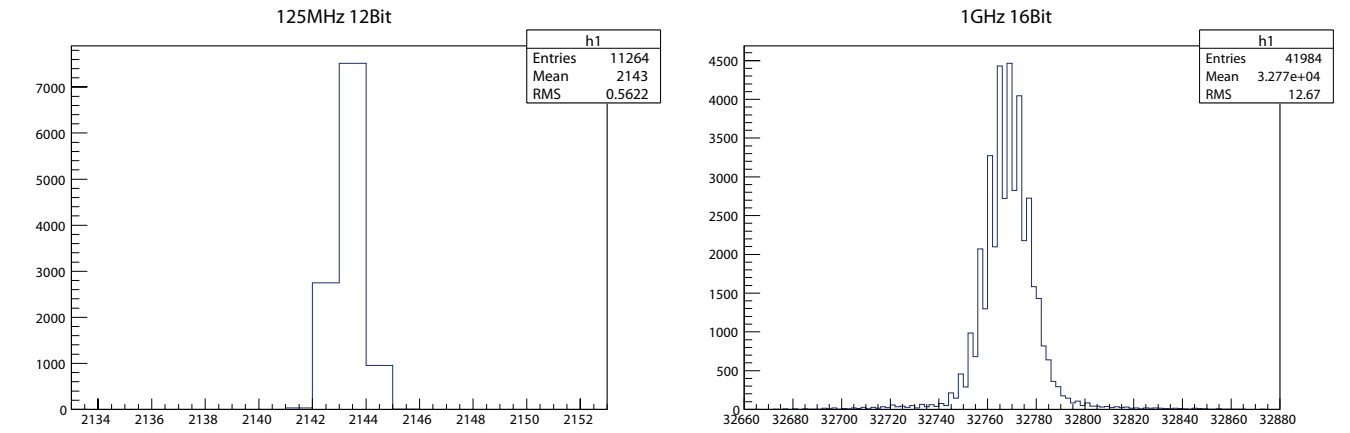

Fig. 6. ENOB measurement for 125-MHz 12-bit and 1-GHz 16-bit FlashADC.

## IV. SUMMARY

The multi-host type DAQ front-end system is proposed and a prototype system is developed. The performance of the FlashADC on the prototype hardware was measured. Further developments such as the trigger management are now in progress.